\documentclass[aps,pra,superscriptaddress,twocolumn,showpacs,titlepage=false]{revtex4-1}
\usepackage{graphicx}
\usepackage{amsmath}
\usepackage{mathrsfs}
\usepackage{dsfont}
\usepackage{amssymb}
\usepackage{subfigure}
\usepackage{color}
\usepackage{bbm}
\usepackage{blkarray}
\usepackage[breaklinks=true,colorlinks=true,linkcolor=blue,urlcolor=blue,citecolor=blue]{hyperref}
\def\bra#1{{\left\langle #1 \right|}}
\def\ket#1{{\left| #1 \right\rangle}}

\providecommand{\proj}[1]{|#1\rangle\langle#1|}
\providecommand{\ketbra}[2]{|#1\rangle\langle#2|}
\providecommand{\sprod}[2]{\langle#1|#2\rangle}
\providecommand{\tr}[1]{{\sf Tr}\left[#1\right]}
\definecolor{amber}{rgb}{1.0, 0.49, 0.0}
\usepackage[dvipsnames]{xcolor}

\renewcommand{\det}{{\sf det}}
\usepackage{verbatim}
\usepackage{amsthm}
\usepackage{enumerate}
\usepackage{bbm}
\usepackage{hyperref}
\newcommand{\bigzero}{\mbox{\normalfont\Large 0}}
\newcommand{\rvline}{\hspace*{-\arraycolsep}\vline\hspace*{-\arraycolsep}}

\begin{document}

\title{Non-Orthogonal Bases for Quantum Metrology}
\author{Marco G. Genoni}
\email{marco.genoni@fisica.unimi.it}
\affiliation{Quantum Technology Lab, Dipartimento di Fisica ``Aldo Pontremoli'', Universit\`a degli Studi di Milano, 20133, Milan, Italy}
\author{Tommaso Tufarelli}
\email{tommaso.tufarelli@nottingham.ac.uk}
\affiliation{School of Mathematical Sciences and Centre for the Mathematics and Theoretical Physics of quantum Non-Equilibrium Systems,
	University of Nottingham, University Park Campus, Nottingham NG7 2RD, United Kingdom}
\date{\today}
\begin{abstract}
{\noindent}Many quantum statistical models are most conveniently formulated in terms of non-orthonormal bases. This is the case, for example, when mixtures and superpositions of coherent states are involved. In these instances, we show that the analytical evaluation of the quantum Fisher information may be greatly simplified by bypassing both the diagonalization of the density matrix and the orthogonalization of the basis. The key ingredient in our method is the Gramian matrix (i.e. the matrix of scalar products between basis elements), which may be interpreted as a metric tensor for index contraction. As an application, we derive novel analytical results for several estimation problems involving noisy Schr\"odinger cat states.
\end{abstract}
\maketitle

\section{Introduction}
{\noindent}The quantum Fisher information (QFI) matrix, and the associated symmetric logarithmic derivatives (SLDs) are central theoretical tools in quantum estimation theory and quantum metrology \cite{helstrom1976quantum,braunstein1994statistical,paris2009quantum,giovannetti2011advances}. Under certain regularity assumptions, the QFI matrix encodes the ultimate precision bounds on the estimation of unknown parameters encoded in a density matrix (know as quantum Cramer-Rao bounds), while the SLDs and their commutators determine whether such bounds may be saturated with physically realizable measurements \cite{ragy2016compatibility,szczykulska2016multi}. The associated applications are plenty, including phase and frequency estimation  \cite{giovannetti2011advances,demkovicz2015interferometry, braun2018RMP, monras2006optimal, escher2011NatPhys,demkovicz2012NatComm, genoni2011prl,chaves2013prl, smirne2016prl, albarelli2017njp, albarelli2018quantum,demkovicze2013gw}, estimation of noise parameters \cite{monras2007loss, pirandola2017noise, bina2018pra,benedetti2018pra,sehdaran2018entropy,brandford2019diffusion}, joint estimation of unitary and/or noisy parameters \cite{ballester2004entanglement, genoni2013dispest, vaneph2013qmqm, humphreys2013quantum, vidrighin2014natcomm, gagatsos2016pra, pezze2017optimal, roccia2017entangling}, sub-wavelength resolution of optical sources \cite{tsang2016quantum,nair2016interferometric,nair2016far,lupo2016ultimate,rehavcek2017multiparameter,YU1,NapoliSuperres}, nano-scale thermometry \cite{stace2010quantum,brunelli2011pra,brunelli2012pra, correa2015individual,mehboudi2019using,mehboudi2018thermometry,campbell2018qsl}, and estimation of Hamiltonian parameters in the presence of phase-transitions \cite{zanardi2008pra, invernizzi2008pra,rossi2017qst}.
The most common approach for calculating the QFI matrix involves the diagonalisation of the density matrix $\rho$ \cite{paris2009quantum}. This method has the tendency to yield tedious calculations, often preventing analytical treatment even in apparently simple quantum statistical models. In continuous-variable systems the calculation can be greatly simplified for Gaussian states, in which case all the quantities of interest are univocally determined by the first and second moments of an appropriate set of quadrature (or canonical) operators \cite{pinel2013pra,monras2013arxiv,gao2014epjd,jiang2014pra, safranek2015njp, serafozzi2017bucca,nichols2018pra}. 
For the most general scenario, Safranek has recently pointed out that the diagonalization of $\rho$ is unnecessary \cite{Safra2018}, and one may build a formal solution for the SLDs via vectorization techniques. However, this approach still assumes the availability of an orthonormal basis spanning the support of $\rho$ and its derivatives. As can be appreciated from several recent works on sub-wavelength resolution, however, the construction of such orthonormal bases can become extremely tedious even for rank-2 states, whenever additional basis elements are needed to span the support of the SLDs  --- for example see Refs.~\cite{tsang2016quantum, YU1} among many others. The purpose of this work is to show that also the construction of an orthonormal basis may be safely bypassed in quantum metrology calculations, by introducing a nontrivial metric tensor for index contraction (the latter is simply the Gramian matrix of the chosen basis). This approach may streamline the calculation of the SLDs and QFI matrix in many cases of interest. We note that a problem-specific version of the theory presented here, which however was not formulated in terms of a modified metric, was already employed by one of the authors in a recent paper \cite{NapoliSuperres}. In the present manuscript, we expand on these ideas to show a fully general and more elegant construction, which may be applied to any finite-rank quantum statistical model formulated in terms of non-orthogonal quantum states. To demonstrate the power of the approach, we provide novel analytical results for several quantum estimation problems involving noisy superpositions of coherent states (or `noisy cat states').

The paper is organised as follows: In Section II we briefly review the quantum Cramer Rao bound and outline the central mathematical problem of quantum estimation theory. i.e. the calculation of the SLDs; In Sec. III we explain the core of our calculation method relying on non-orthonormal bases; in Sec. IV we show examples of novel analytical results that can be obtained with our method, while in Sec. V we draw our conclusions.
\section{quantum Cramer-Rao bounds: brief review}
Following standard local quantum estimation theory, we consider a quantum statistical model of the form
\begin{align}
\rho\equiv\rho(\boldsymbol{\lambda}),
\end{align}
that is, a family of quantum states depending on a set of parameters $\boldsymbol{\lambda}=(\lambda_1,...,\lambda_N)$. It is assumed that the functional form of $\rho$ is known, but that the the exact values of the parameters $\boldsymbol{\lambda}$ are not. The aim of local estimation theory is to set the ultimate bounds on how precisely these parameters can be estimated, given $M\gg1$ copies of the state $\rho$, and assuming the ability to perform any measurement allowed by quantum mechanics. For any \textit{unbiased estimator} $ {\boldsymbol{\lambda}}$, i.e. a function of the measured data providing the correct guess on average, $\mathbb{E}[ {\boldsymbol{\lambda}}]=\boldsymbol{\lambda}$, the quantum Cramer Rao bound (QCRB) holds
\begin{align}
\mathbf{V} \geq \frac{1}{M} \mathbf H^{-1} \,, \label{eq:QCRB}
\end{align}
where $\mathbf V$ is the covariance matrix of the estimator, with elements $V_{\mu\nu}=\mathbb{E}[\lambda_\mu \lambda_\nu]-\mathbb{E}[\lambda_\mu]\mathbb E[ \lambda_\nu]$, while $\mathbf{H}$ is the QFI matrix whose elements are
\begin{align}
H_{\mu\nu}=\frac{1}{2}\tr{\rho \{L_\mu,L_\nu\}}.\label{eq:QFImatrix}
\end{align}
The operators $ L_\mu$ appearing above are known as symmetric logarithmic derivatives (SLDs), and can be found by solving the operator Lyapunov equation
\begin{align}
 2\partial_\mu\rho=\rho L_\mu+ L_\mu\rho,\label{eq:problema}
 \end{align}
 where the shorthand $\partial_\mu\equiv\frac{\partial}{\partial\lambda_\mu}$ shall be used throughout the manuscript. In the multi-parameter case, the matrix inequality (\ref{eq:QCRB}) describing the QCRB is often rewritten in terms of the scalar QCRB 
\begin{align}
\tr{\mathbf{G}\mathbf{V}}\geq\frac{1}{M}\tr{\mathbf G\mathbf H^{-1}},\label{eq:scalarQCRB}
\end{align}
where $\mathbf{G}$ is a generic positive definite matrix (or weight matrix). It is known that the scalar QCRB may be saturated by a physical measurement iff the following matrix elements vanish for all $\mu\neq\nu$:
\begin{align}
\Gamma_{\mu\nu}=\frac{1}{2i}\tr{\rho[L_\mu,L_\nu]}.\label{eq:Gamma}
\end{align}
It is important to note that, in general, saturating the QCRB requires global measurements over many copies of the state $\rho$ \cite{ragy2016compatibility}. However, if the stronger compatibility condition $[L_{\bar \mu},L_{\bar \nu}]=0$ holds for a given pair of indices $\bar{\mu},\bar {\nu}$, the common eigenbasis of $L_{\bar \mu}$ and $L_{\bar \nu}$ provides a projective measurement that can be implemented on a single copy of $\rho$ and is optimal for both $\lambda_{\bar \mu}$ and $\lambda_{\bar \nu}$. It is worth pointing out that Eq.~\eqref{eq:problema} does not have a unique solution whenever the support of $\partial_\mu \rho$ is not fully contained in the support of $\rho$. However, the quantities in Eqs.~\eqref{eq:Gamma} and \eqref{eq:QFI_formula} are uniquely defined.

To conclude this section, we emphasize that solving Eq.~\eqref{eq:problema} embodies the central mathematical problem of quantum estimation theory. In what follows, we outline how Eq.~\eqref{eq:problema} may be converted into a Lyapunov matrix equation by selecting a suitable (not necessarily orthonormal) set of quantum states spanning the support of $\rho$ and its derivatives. 
\section{Expanding operators on a non-orthogonal basis}\label{sec:method}
\label{s:method}
The observation at the core of our work is that the density matrix $\rho$ may be expanded as 
\begin{align}
	\rho=\sum_{ij}R_{ij}\ket{\psi_i}\bra{\psi_j},\label{eq:expansion}
\end{align}
where it is not necessary to assume that the set $\{\ket{\psi_i}\}_i$ is orthonormal - the only requirement is that it should be linearly independent and span the support of $\rho$ and its derivatives. Note that both the coefficients $R_{ij}$ and the states $\ket{\psi_i}$ may depend on the unknown parameters $\boldsymbol{\lambda}$. For brevity we shall refer to the set ${\mathcal B}\equiv\{\ket{\psi_i}\}_i$ as a \textit{basis}, keeping in mind that in general it may span only a subspace of the full Hilbert space.
While the quantum states $\ket{\psi_i}$ may be chosen in whichever way simplifies the problem at hand, it is worth mentioning at least one explicit method to construct such a set. This is comprised of the following steps:
\begin{enumerate}
\item Pick a convenient set of quantum states $\mathcal{B}_{\rho}$, spanning the support of $\rho$ alone. 
\item Compute the set of all the nonzero derivatives \\ $\mathcal{B}'\equiv\big\{\partial_\mu \ket\psi \big\vert\ket{\psi}\in\mathcal{B}_{\rho},\mu \text{ s.t. }\partial_\mu \ket\psi \neq0 \big\}$
\item As a starting point, set $\mathcal{B}=\mathcal{B}_{\rho}$. 
\item For each $\ket{\psi}\in\mathcal {B}'$, check whether $\mathcal{B}\cup\{\ket\psi \}$ is a linearly independent set. If so, enlarge the set $\mathcal B$ by including the state $\ket \psi$; if not, discard $\ket \psi$. 
\end{enumerate}
We note that the above method returns a basis $\mathcal B$ that may be used to expand all the SLDs. In subsection \ref{sec:manybases}, we discuss how the calculation of each $L_\mu$ may be simplified even further by employing a smaller subset $\mathcal B^{(\mu)}\subseteq\mathcal B$, with explicit examples being given in section \ref{sec:badcat}.
In the next subsection we explain how to convert Eq.~\eqref{eq:problema} into a set of matrix Lyapunov equations by introducing an appropriate metric tensor characterizing the set $\mathcal B$.
\subsection{Setting up the SLD equations}
We shall adopt the following convention: a matrix with elements $A_{ij}$ will be denoted as $\mathbf A$. For example, according to Eq.~\eqref{eq:expansion} the state $\rho$ is represented by the matrix $\mathbf R$. Our first step towards solving Eq.~\eqref{eq:problema} is to expand the derivatives of $\rho$ as per 
\begin{equation}
	\partial_\mu \rho=\sum_{ij}D^{(\mu)}_{ij}\ketbra{\psi_i}{\psi_j},
\end{equation}
which defines the matrices $\mathbf D^{(\mu)}$, and we do the same for the (yet unknown) SLDs,
\begin{align}
 L_\mu =\sum_{ij}L^{(\mu)}_{ij}\ketbra{\psi_i}{\psi_j},
\end{align}
implicitly defining the matrices $\mathbf L^{(\mu)}$. The next ingredient we need is the matrix $\mathbf S$ of scalar products between the chosen basis vectors, also known as Gramian matrix. Its elements are
\begin{equation}
S_{ij}\equiv\sprod{\psi_i}{\psi_j}.
\end{equation}
In passing, we remark that the set $\mathcal B$ is linearly independent iff $\det[\mathbf{S}]\neq0$; this can provide a useful sanity check before the start of the procedures outlined below. It can be checked by direct calculation that the matrix $\mathbf S$ acts as a metric tensor for contracting indices. E.g.:
\begin{align}
\rho L_{\mu}&=\sum_{ij}R_{ij}\ket{\psi_i}\bra{\psi_j}\sum_{kl}L_{kl}^{(\mu)}\ketbra{\psi_k}{\psi_l}\nonumber\\
&=\sum_{i,l}\left(\sum_{j,k}R_{ij}L_{kl}^{(\mu)}\underbrace{\sprod{\psi_j}{\psi_k}}_{=S_{jk}}\right)\ketbra{\psi_i}{\psi_l}\nonumber\\
&=\sum_{i,l}\left(\sum_{j,k}R_{ij}S_{jk}L_{kl}^{(\mu)}\right)\ketbra{\psi_i}{\psi_l},
\end{align}
showing that $\rho L_\mu$ corresponds to the matrix $\mathbf R\mathbf S\mathbf L^{(\mu)}$. With a similar calculation, one can show that for example  ${\sf Tr}[\rho L_\mu]= {\sf Tr}[\mathbf R\mathbf S\mathbf L^{(\mu)}\mathbf S]$, where one of the $\mathbf S$ matrices comes from the operator multiplication, the other from the index contraction implicit in the trace. Putting the above steps together, Eq.~\eqref{eq:problema} can be recast as the matrix Lyapunov equation
\begin{align}
	2\mathbf D^{(\mu)}=\mathbf L^{(\mu)}\mathbf S\mathbf R+\mathbf R\mathbf S\mathbf L^{(\mu)},\label{eq:Lyappo}
\end{align}
which may be solved with standard linear algebraic methods to provide the matrices $\mathbf L^{(\mu)}$  and hence the SLDs. This can be achieved either by `brute force', i.e. interpreting Eq.~\eqref{eq:Lyappo} as a linear system of equations for the variables $L^{(\mu)}_{ij}$, or by more sophisticated methods such as matrix vectorization \cite{Safra2018}. Once the SLDs are known, the QFI matrix $H_{\mu\nu}$, as well as the matrix of averaged commutators $\Gamma_{\mu\nu}$ (both real matrices by construction), may be calculated via the relation
\begin{align}
 H_{\mu\nu}+i\Gamma_{\mu\nu}={\sf Tr}[\mathbf R\mathbf S\mathbf L^{(\mu)}\mathbf S\mathbf L^{(\nu)}\mathbf S],
 \label{eq:QFI_formula}
\end{align}
which follows directly from Eqs.~\eqref{eq:QFImatrix} and \eqref{eq:Gamma}. 

 To conclude this section, we explain how the matrices of interest can be in general expressed in terms of standard quantum brakets and the $\mathbf S$ matrix.  Given a generic operator $ A=\sum_{ij}(A)_{ij}\ketbra{\psi_i}{\psi_j}$, we indicate as $\tilde{\mathbf{A}}$ the matrix with elements $(\tilde A)_{kl}=\bra{\psi_k} A\ket{\psi_l}$. With straightforward algebra we get 
\begin{align}
\tilde{\mathbf{A}}=\mathbf{S A S},
\end{align}
hence
\begin{align}
	\mathbf{A}=\mathbf S^{-1}\tilde{\mathbf{A}}\mathbf S^{-1}.
\end{align}

\subsection{Constructing different bases for different parameters}\label{sec:manybases}
So far we have assumed that the full set $\mathcal B$ was employed to calculate each SLD. In practice, to find a particular SLD $ L_\mu$, it is sufficient (and advisable!) to include in Eq.~\eqref{eq:expansion}, and in the subsequent derivations, a subset ${\mathcal B}^{(\mu)}\subseteq{\mathcal B}$ spanning the support of $\rho$ and $\partial_\mu \rho$ only. This in turn defines a reduced Gramian matrix which we will denote as $\mathbf S^{(\mu)}$, while the state $\rho$ will be represented by the matrix $\mathbf R^{(\mu)}$. To avoid making our notation too cumbersome the matrix representation of the SLDs will be indicated by the symbol $\mathbf L^{(\mu)}$ also in this case. Furthermore, in many cases it may not be convenient to construct the set $\mathcal{B}$ by following the procedure outlined at the start of section~\ref{sec:method}, and one may instead begin by constructing each set $\mathcal B^{(\mu)}$ separately. In the simplest scenario, the union $\bigcup_{\mu}\mathcal B^{(\mu)}$ will be linearly independent and can be identified with the set $\mathcal B$ --- in passing we recall that the validity of such an assumption can be checked a posteriori from the condition $\det[\mathbf S]\neq0$. If so, the matrix representations of $\rho$ and $ L_\mu$, with respect to the full basis set ${\mathcal B}$, can be easily found by adding rows and columns of zeros wherever appropriate --- see section \ref{sec:badcat} for an explicit example. In the most general scenario, however, some additional manipulations may be needed in order to construct the set $\mathcal{B}$ by discarding enough redundant (linearly dependent) vectors from the union set $\bigcup_{\mu}\mathcal B^{(\mu)}$. As a consequence, some additional (but straightforward) algebra will be required to find the matrix representations of the SLDs $ L_\mu$ with respect to the full set $\mathcal B$.

In the next section, we shall apply the developed machinery to estimation problems involving coherent states. The latter embody one of the most common and useful examples of a non-orthogonal set of states.
\section{Application: noisy Schr\"odinger cat states}\label{sec:badcat}
A Schr\"odinger cat state is typically defined as the superposition of two coherent states of a single Bosonic mode (or a single quantum harmonic oscillator), that is
\begin{align}
|\psi\rangle_{\sf cat} = \frac{|\alpha\rangle + |-\alpha\rangle}{\sqrt{2 ( 1+ s)}} \label{eq:purecat}
\end{align}
where $s=\sprod{\alpha}{-\alpha}=e^{-2|\alpha|^2}$. The coherent states $|\alpha\rangle$ are defined as the eigenstates of the annihilation operator $a$, i.e. $a|\alpha\rangle = \alpha |\alpha\rangle$, where the Bosonic commutation relation $[a , a^\dag ] =\mathbbm{1}$ is assumed. Since coherent states are traditionally seen as {\em classical states} in quantum optics, the superposition of two coherent states with opposite amplitude is often considered an example of a Schr\"ordinger cat state (i.e. the superposition of two distinct classical configurations). Besides being interesting from a fundamental point of view, Schr\"odinger cat states are considered a resource for quantum information processing \cite{munro2002pra,ralph2003pra,gilchrist2004job,mirrahimi2014njp,vlastakis2013science,leghtas2015science}; in particular we will show in the following how their non-classicality allows to obtain a quantum enhanced precision in the estimation of a displacement in phase space.\\
We will apply the methods discussed in the previous section to an ``imperfect" realisation of Eq.~\eqref{eq:purecat}, i.e. we shall consider states of the form
\begin{align}
\varrho_{\sf cat} &= \mathcal{N} \left[\proj{\alpha}+\proj{-\alpha}+ \right. \nonumber \\
  &\,\,\,\,\,\,\,   \left. c\Big(\ketbra{\alpha}{-\alpha}+\ketbra{-\alpha}{\alpha}\Big)\right],
\label{eq:decoh_cat}
\end{align}
where we have introduced an additional parameter $0\le c\le 1$, which loosely speaking quantifies the coherence of the superposition, and we have defined a normalization constant $\mathcal{N}=1/(2(1+sc))$. By fixing $c=1$ one obtains the pure Schr\"odinger cat state previously introduced, while for $0\leq c<1$, the quantum states are mixed and can be interpreted as ``noisy" (or more precisely partially decohered) Schr\"odinger cat states. For example, if one considers an initial pure cat state with coherent state amplitude $\alpha_0$, evolving in time according to an amplitude damping master equation
\begin{align}
\dot{\varrho} = \gamma a \varrho a^\dag - \gamma \left( a^\dag a \varrho + \varrho a^\dag a \right)/2 \,,
\end{align}
one obtains that the evolved quantum state at time $t$ has the form as in Eq. (\ref{eq:decoh_cat}) with parameters
\begin{align}
\alpha &= \alpha_0 e^{- \gamma t /2} \, , \nonumber \\
c &= \exp\left\{- 2 |\alpha_0|^2 ( 1- e^{-\gamma t} )\right\} \,. \label{eq:lossyparameters}
\end{align}
In the following we will first discuss the estimation of the parameters $\boldsymbol{\lambda} = \{ c , \alpha \}$ characterizing the quantum state $\varrho_{\sf cat}$, and via a simple change of variables, the parameters $\tilde{\boldsymbol{\lambda}} = \{ \bar\gamma = \gamma t , \alpha_0 \}$, characterizing the evolution of a pure Schr\"odinger cat state in a lossy channel (notice that in the following, for the sake of simplicity, we will consider $\alpha \in \mathbbm{R}$ and $\alpha_0 \in \mathbbm{R}$). Next, we will consider this class of states as probes for the estimation of a unitary displacement in phase space, discussing in detail the robustness of the estimation precision against decoherence, and comparing the results with the ones obtainable with lossy squeezed states as inputs.
\subsection{Estimation of parameters characterizing a noisy cat state}
Following the methods in Sec.~\ref{s:method}, we will initially discuss the estimation of the two parameters $\boldsymbol{\lambda} = \{ c , \alpha \}$ separately, i.e. we shall calculate the diagonal QFI matrix elements $H_{cc}$ and $H_{\alpha\alpha}$. Then, we will show how to calculate the remaining element $H_{c\alpha}$  and hence the full quantum Fisher information matrix for the two parameters, as well as the quantity $\Gamma_{c\alpha}$ characterizing the joint estimation properties. Finally, we shall extend the discussion to the parameters $\tilde{\boldsymbol{\lambda}} = \{ \bar\gamma = \gamma t , \alpha_0 \}$.\\
For the {\em decoherence} parameter $c$ it is straightforward to observe that we can describe all the operators in terms of the two-dimensional basis $\mathcal{B}^{(c)} = \{ |\alpha\rangle , |-\alpha\rangle \}$ only. We can immediately identify the matrices of interest:
\begin{align}
\mathbf{R}^{(c)}&=\mathcal{N}\begin{pmatrix}
1&c\\
c&1
\end{pmatrix} \,,\\
\mathbf{S}^{(c)}&=\begin{pmatrix}
1& s\\
 s&1
\end{pmatrix} \,,\\
\mathbf{D}^{(c)}&=\mathcal{N}\begin{pmatrix}
0& 1\\
1&0
\end{pmatrix}- (2 s \mathcal{N}) \, \mathbf{R}^{(c)},
\end{align}
where the last expression follows from
\begin{equation}
\partial_c\varrho =\mathcal{N}\Big(\ketbra{\alpha}{-\alpha}+\ketbra{-\alpha}{\alpha}\Big)- (2 s \mathcal{N}) \, \varrho \,.
\end{equation}
The Lyapunov equation $2\mathbf D^{(c)}=\mathbf L^{(c)}\mathbf S^{(c)}\mathbf R^{(c)}+\mathbf R^{(c)}\mathbf S^{(c)}\mathbf L^{(c)}$ can be solved analytically, yielding
\begin{equation}
	\mathbf L^{(c)}=\mathcal{N} \left(
	\begin{array}{cc}
	-2 \frac{c s^2+2 s+c}{\left(1 - c^2\right) \left(1-s^2\right)} & 2 \frac{s^2+2 c s+1}{\left(1-c^2\right) \left(1-s^2\right)} \\
	2 \frac{s^2+2 c s+1}{\left(1-c^2\right)  \left(1-s^2\right)} & - 2 \frac{c s^2+2 s+c}{\left(1- c^2\right) \left(1-s^2\right)} \\
	\end{array}
	\right)
\end{equation}
We can then calculate the QFI via Eq.~(\ref{eq:QFI_formula}) obtaining
\begin{align}
	H_{c,c}&={\sf Tr}[\mathbf{R}^{(c)}\mathbf{S}^{(c)}\mathbf{L}^{(c)}\mathbf{S}^{(c)}\mathbf{L}^{(c)}\mathbf{S}^{(c)}] \nonumber \\
	&=\frac{1-e^{-4 \alpha^2}}{\left(1-c^2\right) (e^{-2 \alpha^2} c+1)^2} \,.
\end{align}
The QFI is monotonically increasing with the amplitude $|\alpha|$, and in particular in the limit of large coherent amplitudes $|\alpha|\to\infty$ we have $H\to(1-c^2)^{-1}$ (note that the QFI presents a divergence at $c=1$, where the state changes rank and the quantum Cramer Rao bound does not hold \cite{SafranekDiscQFI,SevesoDiscQFI}).\\

If we now focus on the estimation of the parameter $\alpha$, we find that we can expand the operators of interest in the extended basis $\mathcal{B}^{(\alpha)}=\{|\alpha\rangle, |-\alpha\rangle, \partial_\alpha |\alpha\rangle, \partial_\alpha |-\alpha\rangle \}$. By observing that $\partial_\alpha |\alpha\rangle = (a^\dag-a) |\alpha\rangle$, one can evaluate the scalar product matrix as
\begin{align}
\mathbf{S}^{(\alpha)}&=\begin{pmatrix}
1 & s & 0 & - 2 s \alpha \\
 s & 1 & -2 s \alpha & 0 \\
 0 & -2 s \alpha & 1 & s (4 \alpha^2 -1) \\
 -2 s \alpha & 0 & s (4 \alpha^2 -1 ) & 1
\end{pmatrix} \,.
\end{align}
As anticipated in section ~\ref{sec:method}, the state matrix in terms of the new basis can be trivially found by adding appropriate rows and columns of zeros, i.e.
\begin{align}
\mathbf{R}^{(\alpha)}&=\mathcal{N} 
\begin{pmatrix}
  \begin{matrix}
  1 & c \\
  c & 1
  \end{matrix}
  & \rvline & \bigzero \\
\hline
  \bigzero & \rvline &
 \bigzero
\end{pmatrix} \,,
\end{align}
while differentiating $\varrho_{\sf cat}$ with respect to the parameter $\alpha$, one obtains
\begin{align}
\mathbf{D}^{(\alpha)}&=\mathcal{N} 
\begin{pmatrix}
\bigzero & \rvline &
\begin{matrix}
  1 & c \\
  c & 1
  \end{matrix} \\
\hline
\begin{matrix}
  1 & c \\
  c & 1
  \end{matrix}
& \rvline &
 \bigzero
\end{pmatrix} + \frac{4 \alpha s c}{1+ sc} \,\mathbf{R}^{(\alpha)} \,.
\end{align}
Also in this case the Lyapunov equation for $\mathbf L^{(\alpha)}$ can be solved analytically; we do not report the full solution here as the formula is cumbersome and does not give any particular insight. However it is important to notice that, as the support of $\rho_{\sf cat}$ is smaller than the support of $\partial_\alpha\rho_{\sf cat}$, the SLD operator $L_\alpha$, and as a consequence the matrix $\mathbf L^{(\alpha)},$ is not unique. The corresponding QFI (which is instead uniquely defined) has a much simpler analytical formula:
\begin{align}
	H_{\alpha,\alpha}&={\sf Tr}[\mathbf{R}^{(\alpha)}\mathbf{S}^{(\alpha)}\mathbf{L}^{(\alpha)}\mathbf{S}^{(\alpha)}\mathbf{L}^{(\alpha)}\mathbf{S}^{(\alpha)}] \nonumber \\
	&=\frac{4 \left( 1 - c^2 e^{-4 \alpha^2} +4 \,c\, \alpha^2 e^{-2 \alpha^2} \right)}{(1 + c\, e^{-2 \alpha^2})^2} \,.
\end{align}
By inspecting this formula one observes a non trivial behaviour both in terms of the coherent states amplitude $\alpha$ and of the coherence parameter $c$. \\

In the case we are considering here, the basis $\mathcal{B}^{(c)}$ is a strict subset of the basis $\mathcal{B}^{(\alpha)}$. As ${\sf det}[\mathbf{S}^{(\alpha)}] \neq 0$, one can simply extend the matrix $\mathbf L^{(c)}$ with block matrices filled with zeros, and directly use the matrices $\mathbf{R}=\mathbf{R}^{(\alpha)}$ and $\mathbf{S}=\mathbf{S}^{(\alpha)}$ for all the calculations. In particular one obtains a non-zero off-diagonal element by using the formula
\begin{align}
	H_{c,\alpha}&=\frac12 \left( {\sf Tr}[\mathbf{R}\mathbf{S}\mathbf{L}^{(c)}\mathbf{S}\mathbf{L}^{(\alpha)}\mathbf{S}] \right.  \nonumber \\
	& \,\,\,\,\,\,\,\, \left. + {\sf Tr}[\mathbf{R}\mathbf{S}\mathbf{L}^{(\alpha)}\mathbf{S}\mathbf{L}^{(c)}\mathbf{S}] \right)
	\nonumber \\
	&=-\frac{4 \alpha e^{-2 \alpha^2}}{( 1+ce^{-2 \alpha^2})^2} \,,
\end{align}
showing that the estimation of the two parameters is ultimately correlated.\\
The same matrices can be exploited to investigate the joint estimation properties of the two parameters. In particular one can calculate the average value of the commutator between the SLD operators,
\begin{align}
{\sf Tr}[\varrho_{\sf cat} [ {L}_c , {L}_\alpha ] ] &={\sf Tr}[\mathbf{R}\mathbf{S}\mathbf{L}^{(c)}\mathbf{S}\mathbf{L}^{(\alpha)}\mathbf{S}]
	- {\sf Tr}[\mathbf{R}\mathbf{S}\mathbf{L}^{(\alpha)}\mathbf{S}\mathbf{L}^{(c)}\mathbf{S}]
	\nonumber \\
	&=0  \,.
\end{align}
This {\em weak commutativity condition} ensures that the scalar bound on the variances of the two parameters can be achieved, but in principle this will require a joint measurement on an asymptotically large number of copies of the quantum state $\varrho_{\sf cat}$ \cite{ragy2016compatibility}.\\

One can apply the same machinery if we want to consider the estimation of the parameters $\tilde{\boldsymbol{\lambda}}= \{\bar{\gamma}=\gamma t, \alpha_0\}$ characterizing respectively the loss parameter of the master equation and the initial amplitude of the pure cat state. In fact, by a simple change of variables one can obtain the corresponding SLD operators and QFI matrix \cite{paris2009quantum} as in the equations
\begin{align}
\widetilde{\bf H} &= {\bf B H B}^{\sf T} \,, \\
\mathbf{L}^{(\bar\gamma)} &= B_{11} \mathbf{L}^{(c)} + B_{12} \mathbf{L}^{(\alpha)} \,, \\
\mathbf{L}^{(\alpha_0)} &= B_{21} \mathbf{L}^{(c)} + B_{22} \mathbf{L}^{(\alpha)} \,.
\end{align}
where ${\bf H}$ and $\widetilde{\bf H}$ denote respectively the QFI matrix for the old and new parameters, and where we have introduced the Jacobian matrix with elements $B_{\mu\nu} =  \partial \lambda_\nu / \partial \tilde\lambda_\mu$. As the analytical formula are quite cumbersome and do not give any particular insight we decided to not report them here. By using the new SLD operators we can however prove that the weak-commutativity condition holds also for the new couple of parameters, i.e. ${\sf Tr}[\varrho_{\sf cat} [ {L}_{\bar\gamma} , {L}_{\alpha_0} ] ] = 0 $, showing how $\bar{\gamma}$ and $\alpha_0$ can in principle be jointly estimated.
\subsection{Displacement estimation with noisy Schr\"odinger cat states}
We now consider the (single-parameter) estimation of displacement in phase-space: an initial probe state $\varrho_0$ undergoes the unitary transformation
\begin{align}
\varrho_0 \rightarrow  \varrho_\epsilon = D(i\epsilon) \varrho_0 D(i\epsilon)^\dag \,,
\end{align}
where $D(\beta) = \exp\{\beta a^\dag - \beta^* a\}$ denotes the displacement operation in phase space, that, in the case of a purely immaginary parameter $\beta= i\epsilon$, describes a displacement along the ${p} = i (a^\dag-a)/\sqrt{2}$ quadrature. In the case of a pure coherent state as probe$, \varrho_0 = |\alpha\rangle\langle \alpha|$ the QFI is constant, equal to $H_{|\alpha\rangle} = 4$. A non-classical scaling in terms of the average photon number $\bar n = {\sf Tr}[\varrho_0 a^\dag a]$ can on the other hand be achieved by employing pure non-classical states such as Schr\"odinger cat states as the one in Eq. (\ref{eq:purecat}) \cite{munro2002pra,gilchrist2004job}. In particular, in this case one gets ($\alpha \in \mathbbm{R}$)
\begin{align}
H_{|\psi\rangle_{\sf cat}} = \frac{4 (4 \alpha^2 + 1 + e^{-2 \alpha^2} )}{1 + e^{-2 \alpha^2}} \stackrel{\alpha \gg 1}{\approx}16 \bar n + 4 \,,
\end{align}
where the average photon number reads $\bar n = \alpha^2 (1 - e^{-2 \alpha^2})/(1+e^{-2 \alpha^2})$. A similar linear scaling for large average photon numbers is achieved for input squeezed states \cite{munro2002pra,genoni2009pra}. \\

\begin{figure}[h!]
\center
\includegraphics[width=0.49\textwidth]{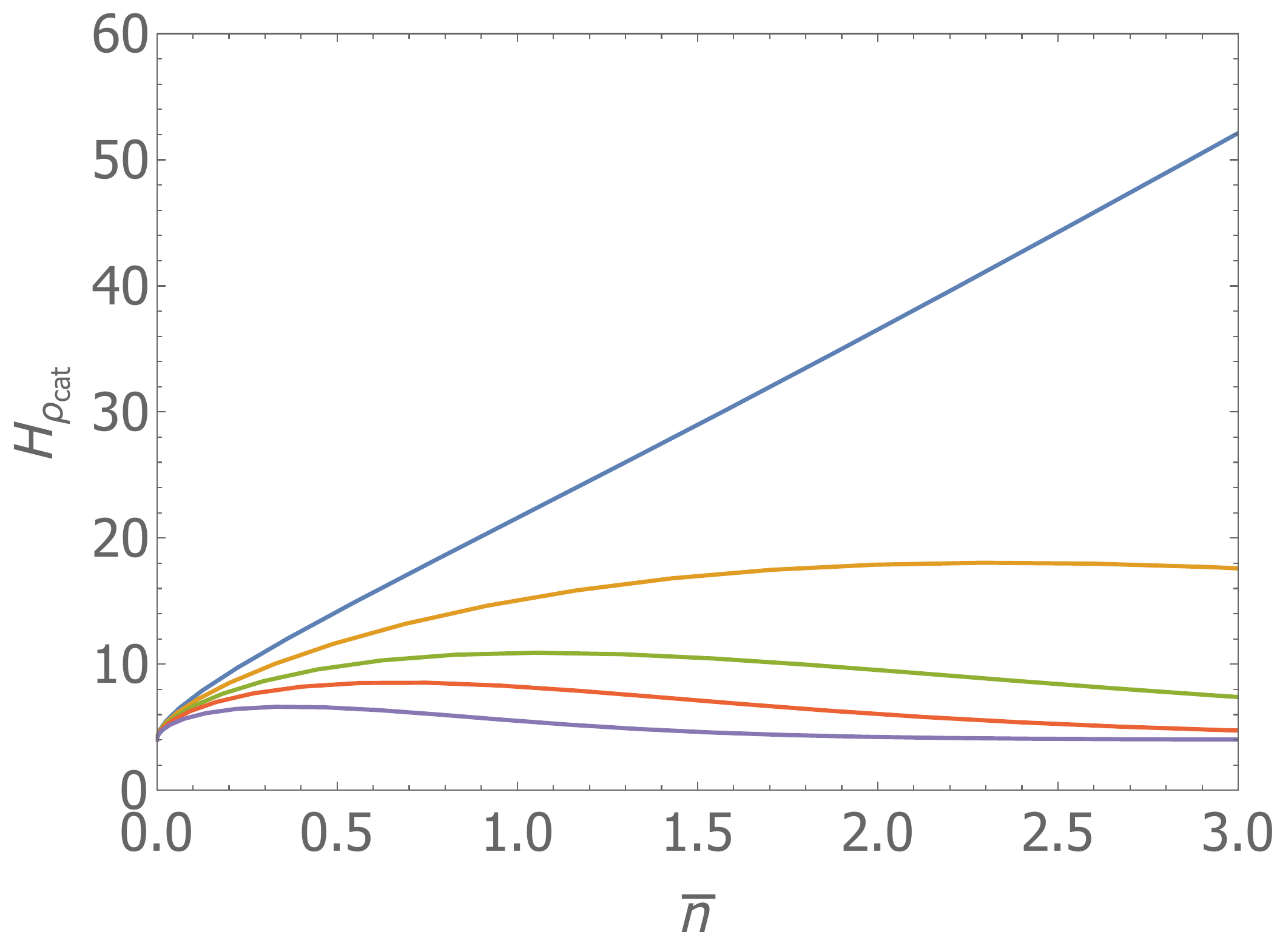} 
\caption{QFI for displacement estimation with input Schr\"odinger cat states $\varrho_{\sf cat}$ evolved in a lossy channel, for fixed $\bar \gamma = \gamma t$, as a function of the average photon number $\bar n = {\sf Tr}[\varrho_{\sf cat} a^\dag a]$. From top to bottom:$ \bar \gamma = \{0, 0.1, 0.2, 0.3, 0.5\}$. }
\label{f:dispest_lossycat}
\end{figure}

We here want to exploit the method discussed in Sec. \ref{s:method}, in order to calculate the QFI for the lossy Schr\"odinger cat state in Eq. (\ref{eq:decoh_cat}) as input $\varrho_0$. In this case one can employ the basis set 
$$
\mathcal{B}^{(\epsilon)} = \{ D(i \epsilon) |\alpha\rangle , \, D(i \epsilon) |-\alpha\rangle , \,\partial_\epsilon D(i \epsilon) |\alpha\rangle, \, \partial_\epsilon D(i \epsilon) |-\alpha\rangle \} \,,
$$ 
leading to the following matrices
\begin{align}
\mathbf{S}^{(\epsilon)}&=\begin{pmatrix}
1 & s & 2 i \alpha & 0  \\
 s & 1 & 0 & -2 i \alpha \\
 -2 i \alpha & 0 & 1 + 4 \alpha^2 & s \\
 0 & 2 i \alpha & s  & 1 + 4 \alpha^2 
\end{pmatrix} \,, \\
\mathbf{R}^{(\epsilon)}&=\mathcal{N} 
\begin{pmatrix}
  \begin{matrix}
  1 & c \\
  c & 1
  \end{matrix}
  & \rvline & \bigzero \\
\hline
  \bigzero & \rvline &
 \bigzero
\end{pmatrix} \,, \\
\mathbf{D}^{(\epsilon)}&=\mathcal{N} 
\begin{pmatrix}
\bigzero & \rvline &
\begin{matrix}
  1 & c \\
  c & 1
  \end{matrix} \\
\hline
\begin{matrix}
  1 & c \\
  c & 1
  \end{matrix}
& \rvline &
 \bigzero
\end{pmatrix} \,.
\end{align}
Remarkably, an analytical solution of the Lyapunov equation can be obtained for the SLD operator, in terms of the matrix
\begin{align}
\mathbf{L}^{(\epsilon)}&=\begin{pmatrix}
0 & -\frac{4 i \alpha}{s(1+sc)} & 0 & \frac{2}{s} \\
\frac{4 i \alpha}{s(1+sc)} & 0 & \frac2s & 0 \\
0 & \frac2s & 0 & \frac{i}{s\alpha} \\
\frac2s & 0 & -\frac{i}{s \alpha} & 0 
\end{pmatrix} \,, 
\end{align}
yielding a QFI 
\begin{align}
H_{\varrho_{\sf cat}} = \frac{4\left[ 4 \alpha^2 \left(c^2 + c\, e^{-2 \alpha^2} \right) + \left(1  + c \,e^{-2 \alpha^2}\right)^2\right]}{(1 +c \,e^{-2 \alpha^2})^2} \,.
\end{align}
As the average photon number of a lossy cat state reads
\begin{align}
\bar{n} = \alpha^2 \left( \frac{1 - c \,e^{-2 \alpha^2}}{1 + c \, e^{-2 \alpha^2} } \right) \,,
\end{align}
by fixing the noisy constant $c$ and varying the average photon number $\bar n$ with the coherent amplitude $\alpha$, one gets  
\begin{align}
H_{\varrho_{\sf cat}} \stackrel{\bar n \gg 1}{\approx}16 c^2 \bar n + 4 \,,
\end{align}
that is a linear scaling in $\bar n$ is observed for any value of $c>0$. A different behaviour is however observed if we substitute the parameters as in Eqs. (\ref{eq:lossyparameters}), and look at the {\em physically more relevant} parameters $\alpha_0$ and $\bar\gamma$, corresponding to a pure Schr\"odinger cat state evolving in a lossy channel. In fact in this case, by fixing $\bar{\gamma}$ and varying the photon number with the initial amplitude $\alpha_0$ one obtains the plot in Fig. \ref{f:dispest_lossycat}: the QFI is not even monotonically increasing with $\bar n$ as soon as $\bar\gamma>0$, and a maximum  $\bar n_{\sf max}$ is observed, whose value monotonically decreases with $\bar\gamma$.\\

\begin{figure}[h!]
\center
\includegraphics[width=0.49\textwidth]{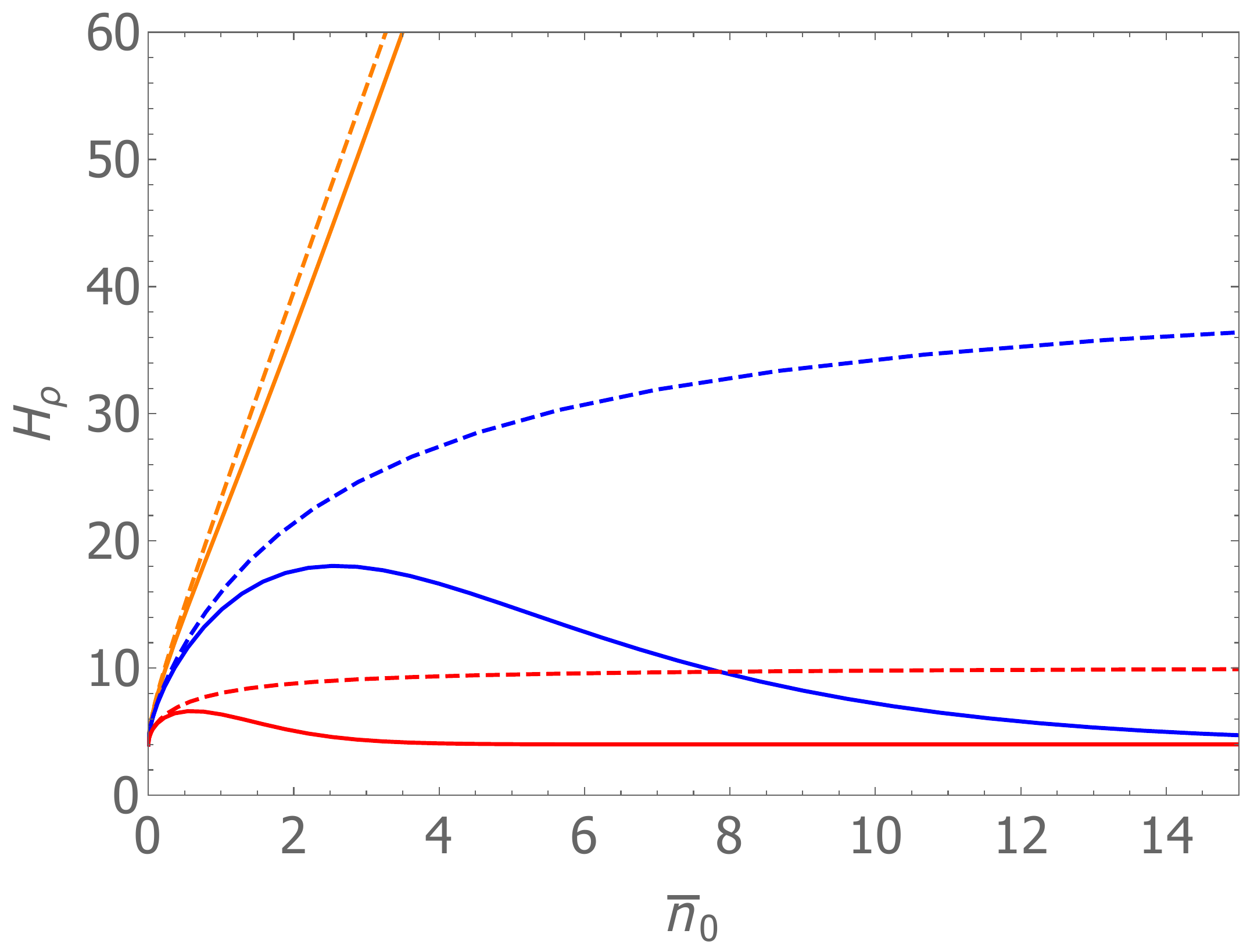} 
\caption{QFI for displacement estimation with respectively input Schr\"odinger cat states $\varrho_{\sf cat}$ (solid lines) and squeezed states (dashed lines)  evolved in a lossy channel, for fixed $\bar \gamma = \gamma t$, as a function of the input pure state average photon number $\bar n_0 = {\sf Tr}[\varrho a^\dag a]$. From top to bottom:$ \bar \gamma = \{0, 0.1, 0.5\}$. 
}
\label{f:robustness}
\end{figure}

In order to investigate the robustness of the estimation properties of cat states, we have also looked at how the QFI varies, at fixed $\bar\gamma$, varying the average photon number of the initial pure cate state, characterized by the coherent states amplitude $\alpha_0$, that enters in the lossy channel. We compare if with the QFI corresponding to a initial pure squeezed state $|\psi\rangle_{\sf sq} = S(r) |0\rangle$, where $S(r) = \exp\{r (a^{\dag 2} - a^2) \}$ is the squeezing unitary operator, characterized by initial photon number $\bar n_0 = {}_{\sf sq}\langle \psi | a^\dag a |\psi\rangle_{\sf sq} = \sinh^2 r$. The corresponding QFI can be indeed evaluated analytically, by exploiting formulas for single-mode Gaussian states \cite{pinel2013pra}, obtaining
\begin{align}
H_{\varrho_{\sf sq}} = \frac{4 e^{2 r}}{e^{2 r}(1-e^{-\gamma t})+e^{-\gamma t}} \,.
\end{align}
In Fig. \ref{f:robustness}, we plot both QFIs for a fixed value of $\bar \gamma$ and as a function of the initial photon number $\bar n_0$. In both cases the linear scaling is lost in the presence of noise, but while also in this case, as soon as $\gamma >0$, the cat state's QFI is not monotonous and presents a maximum in $\bar n_0$, the squeezed state QFI is always monotonically increasing and reaches a limiting value. In more detail one finds that, for fixed $\bar \gamma >0$, in the limit of respectively large $\alpha$ and large $r$,
\begin{align}
\lim_{\alpha \rightarrow \infty} H_{\varrho_{\sf cat}} &= 4 \,, \\
\lim_{r \rightarrow \infty} H_{\varrho_{\sf sq}} &= \frac{4}{1-e^{-\gamma t}} \,,
\end{align}
that is, while the cat states eventually yield the coherent state limit, a constant enhancement can be obtained if we exploit input squeezed states.
This result indeed confirms the intuition that Schr\"odinger cat states are very fragile under the loss master equation: as one can observe from Eqs. (\ref{eq:lossyparameters}), the decoherence parameter $c$ goes exponentially to zero, with an exponent that increases with $|\alpha_0|^2$, i.e. the larger is the energy of the cat state, the faster is the decay of the non-classical {\em coherence} terms.
\section{Conclusions}
In summary, we presented a general method to perform quantum metrology calculations when $\rho$ and its derivatives are initially expanded on a non-orthonormal basis. By avoiding two significant computational hurdles, i.e. the diagonalization of the density matrix and the orthogonalization of the basis, our method has the potential to provide novel analytical results in a  variety of quantum estimations problems. While the method is naturally suited for quantum statistical models involving coherent states, we are confident that it will prove useful for a much wider class of models, for example involving superpositions and mixtures of general Gaussian states or spin coherent states among many others.
\section*{Acknowledgments}
We thank G. Adesso, F. Albarelli, D. Dorigoni, C. Napoli, and S. Piano for the useful discussions. 
MGG acknowledges support from a Rita Levi-Montalcini fellowship of MIUR. 
TT acknowledges support from the University
of Nottingham via a Nottingham Research Fellowship.
%
%
\bibliography{Bib_staceppa}
\bibliographystyle{apsrevfixedwithtitles}
\end{document}